# FABRICATION EXPERIENCE OF THE PRE-PRODUCTION PIP-II SSR2 CAVITIES AT FERMILAB*

M. Parise†, D. Passarelli, P. Berrutti, V. Roger, Fermi National Accelerator Laboratory, 60510 Batavia, IL, USA
D. Longuevergne, P. Duchesne, CNRS, Université Paris-Saclay, Université Paris Sud,
IJCLab, 91405 Orsay, France

## Abstract

The Proton Improvement Plan-II (PIP-II, [1]) linac will include 35 Single Spoke Resonators type 2 (SSR2). A total of eight pre-production SSR2 jacketed cavities will be procured and five installed in the first pre-production cryomodule. The mechanical design of the jacketed cavity has been finalized and it will be presented in this paper along with fabrication and processing experience. The importance of interfaces, quality controls and procurement aspects in the design phase will be remarked as well as lessons learned during the fabrication process. Furthermore, development studies will be presented together with other design validation tests.

## INTRODUCTION

The Radio-Frequency (RF) volume final design and the advanced (but not final) mechanical design of the pre-production SSR2 cavities was presented 3 years ago at the 19th International Conference of RF Superconductivity (SRF'19) [2], [3]. An internal project Final Design Review was held in November 2019 and the procurement of niobium initiated shortly thereafter. The procurement of the jacketed cavities, to be manufactured in industry, initiated at the beginning of 2021 with the first bare cavity completed at the end of the same year. After bulk Buffered Chemical Polishing (BCP), High Pressure Rinse (HPR) of the RF volume and High Temperature Heat Treatment (HTHT) the jacketing and room temperature tests of the first unit were completed in July 2022. The next units are expected to be completed soon.

## DESIGN OVERVIEW

Some of the main parameters used for the design of the SSR2 cavities are reported in Table 1. The RF volume is optimized not only to enable the best Electro Magnetic (EM) performance but, especially on this spoke resonator, to mitigate multipacting. The result is the starting point for the mechanical design: a multi-objective optimization problem revolving around not only the structural soundness but also involving RF parameters such as frequency sensitivity and Lorentz Force Detuning (LFD). For this cavity, the design of the bare cavity and the helium jacket is carried out simultaneously as a integrated system opposed to trying to optimize the bare cavity first and only secondly the helium jacket which may have led to unnecessary design iterations. Lessons learned from the previous generation of spoke resonators are also considered carefully as of this process [4], [5]. Fig. 1 shows the finalized design of the cavity and a section defining the helium space and beam volume.

Table 1: Parameters of the SSR2 Cavity Discussed in this Paper

| Parameter | Value |
|---|---|
| Nominal Frequency, MHz | 325 |
| df/dp, $\frac{Hz}{mbar}$ | <25 |
| Target Frequency Allowable Error, kHz | +/-50 |
| Maximum Allowable Working Pressure (MAWP) RT / 2 K, bar | 2.05 / 4.1 |

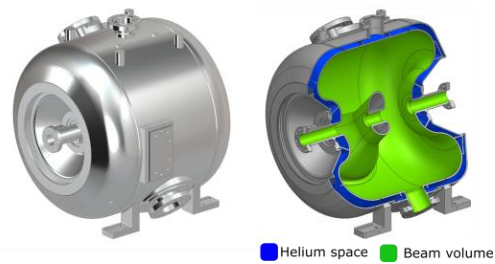

Figure 1: Cavity final design and definition of helium space and beam volume. The section view enables to see the complex geometry of the niobium structure encompassing the beam volume.

### Design Choices with an Impact on the Fabrication

As a result of the multipacting mitigation, the endwalls of the cavity present an elliptical profile with an inflection point and overall high depth to radius ratio making the part difficult to form. Moreover, the structural soundness, in compliance with Fermilab internal guidelines and the ASME Boiler and Pressure Vessel Code [6], whenever possible, required to procure niobium sheets with a minimum thickness close to 5 mm. This value is the maximum used among all superconducting cavities of PIP-II and above the usual thickness that cavity fabricators are used to. Thick material implies a more difficult forming and Electron-Beam (EB) welding operations. To simplify the procurement and reduce the number of joint types, the SSR2 cavities are manufactured only from 2 materials: high purity niobium and titanium. Not using nb-ti alloys means directly joining niobium and titanium through EB and Tungsten Inert Gas (TIG) joints, a practice that has never been used not only for PIP-II but at Fermilab in general.

## BARE CAVITY FABRICATION

Fig. 2 shows the first pre-production SSR2 cavity as fabricated. The cavity mostly consists of niobium sheets that are formed and joined together through EB welding. Titanium reinforcements are also cut from sheets, formed and welded to the niobium.

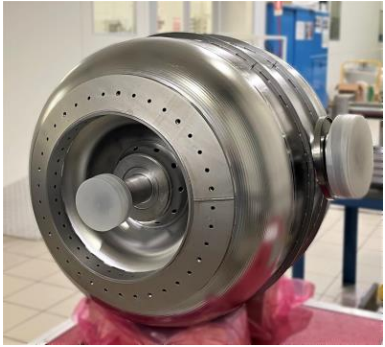

Figure 2: Pre-production SSR2 Bare Cavity.

### Niobium Forming

The most critical components to form are the endwalls, the spokes and the spoke collars. Copper sheets are used in-lieu of the precious niobium to verify that the forming technique and equipment is capable of producing components within the mechanical tolerances. Once the process is verified to produce the expected results, the same steps are put in place to obtain the final components. The endwalls are fabricated using the metal spinning technique while the spokes and spoke collars are obtained using deep drawing.

The most severe thickness reduction takes place on the endwalls. The thickness post spinning is measured at 36 different locations (9 points spaced by 90 degrees as shown in Fig. 3) and the results are reported in Fig. 4 as an average (over 6 endwalls) percentage thickness reduction at each location. The error bars extremities represent the maximum and minimum values over the 6 endwalls. The maximum is 20 % in correspondence of the inflection.

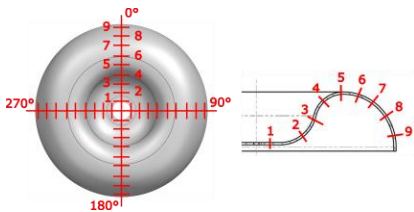

Figure 3: Diagram Showing the Points at which Location the Thickness is Measured.

### Welding

All the bare cavity components are joined using exclusively the EB welding technique for a total of 59 joints that can be divided in 8 different types. The helium vessel is welded together and to the bare cavity using a mix of TIG and

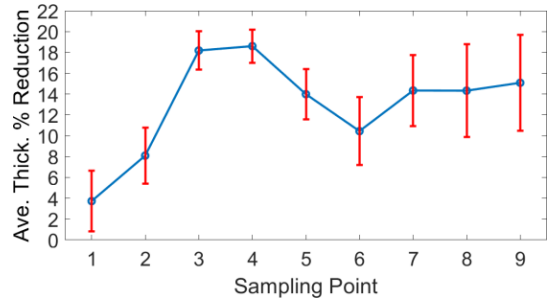

Figure 4: Average Percentage Thickness Reduction on the Endwalls used to Fabricate the first 3 SSR2 Cavities.

EB welding technique (this brings the total type of EB joints to 11 and additional 2 types of TIG welds). The challenges specific of the SSR2 cavities are: the presence of 4 mm thick full penetration niobium-niobium EB joints performed from 1 side only and of niobium to titanium EB and TIG direct joints. Each of the 13 types required a Welding Procedure Specification (WPS) a Procedure Qualification Record (PQR) and a Welder Performance Qualification (WPQ) according to the Code [6] and to Fermilab guidelines. Each joint on each component is also visually inspected and radiographs are taken for the most criticals. Fig. 5 provides some details of the thickest nionbium to niobium EB welded joints face and root and also shows one of the EB welded joints between niobium and titanium. The smoothness and uniformity of the joints' surfaces is essential for the cavity to work properly in presence of EM fields.

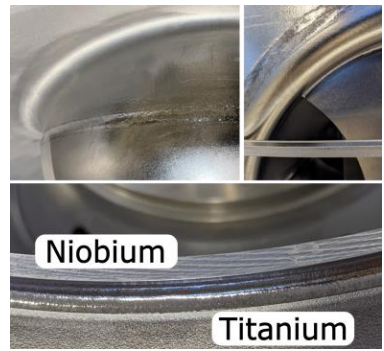

Figure 5: At the top: thick EB welded joints niobium to niobium on the RF and no-RF sides. At the bottom: titanium to niobium EB welded joint.

### Frequency Trimming and Inelastic Tuning

To achieve the required operating frequency of 325 MHz in operation, several target frequencies are established at different step of the cavity fabrication and processing. The targets are based on past experience [4] and on simulation results. Given the operational elastic tuning range that the tuner can provide, each target frequency is allowed to deviate by +/- 50 kHz. If at any manufacturing step, the resonant frequency is outside this range and inelastic tuning is performed. The first manufacturing step during which the resonant frequency can be measured is the so called frequency trimming: the 2 endwall sub-assemblies and the main shell sub-assembly are pre-assembled (see Fig. **??** and the resonant frequency of the cavity is measured. The main shell is therefore trimmed on each side to meet the target frequency and finally the 2 final circumferential EB welds are performed.

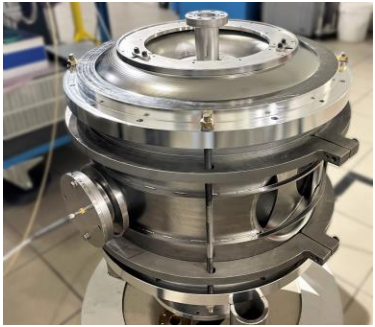

Figure 6: SSR2 Cavity Pre-Assembled Using a Fixture to Perform a Frequency Measurement before the Frequency Trimming of the Main Shell.

Table 2 reports the frequency shifts observed during the frequency trimming operation on the first 3 cavities as well as the first leak check. On the first cavity, in order to reduce the uncertainty of the process, the welding operation is divided into 2 steps: only one endwall is welded to the shell, the cavity is removed from the EB welding machine and the frequency is measured and compared to the simulation results, after the second endwall is welded.

Table 2: Frequency Shifts during Frequency Trimming and Leak Check. Values are in MHz.

| Step | Cav. 1 | Cav. 2 | Cav. 3 |
| --- | --- | --- | --- |
| 1st Trim | -1.396 | -1.418 | -1.306 |
| 2nd Trim | -1.552 | -1.595 | -2.053 |
| 1st EB Weld | -0.135 | - | - |
| Cav. Welded | -0.090 | -0.155 | -0.332 |
| Leak Check | -0.004 | -0.007 | -0.005 |

## JACKETING

After the bare cavity goes through the processing steps the jacketing operation starts. Fig. 7 shows the trend of the resonant frequency at different discrete steps during this process for the first SSR2 cavity as well as the final pressure test and leak check. The greatest frequency shift of about 100 kHz occurs welding the cone Ring End (RE) and the Inner Coupling Ring (see Fig. 8). This welds rigidly connects 2 areas of the bare cavity and this was expected. On the other hand, the weld between the side ports and the helium vessel (another rigid connection between bare cavity and helium vessel) generates a very minor frequency shift due to the clever solution adopted to connect the parts [3].

Table 3 represent the frequency shifts during the procedure. The pressure is gradually incremented until the 117 % of the MAWP at room temperature (2.4 bar-g). At each step the pressure is held for 1 minute and 10 minutes at the step 10 while checking for leaks or pressure drops at the minimum scale of the pressure gauge. This cavity successfully passed the pressure test also showing minimal overall frequency shift.

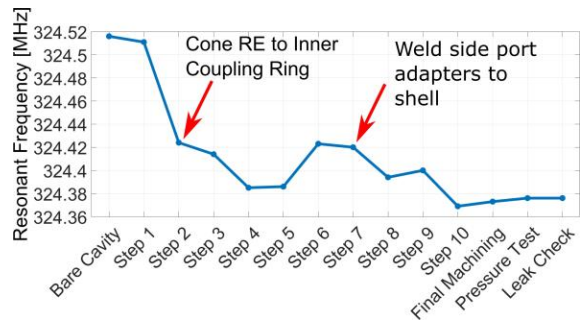

Figure 7: Resonant Frequency Shifts during the Jacketing operation of the first SSR2 Cavity.

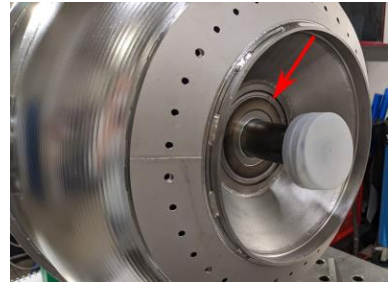

Figure 8: The Weld between the Cone RE and the Inner Coupling Ring is the one Generating the Greatest Frequencty Shift during the Helium Vessel Integration

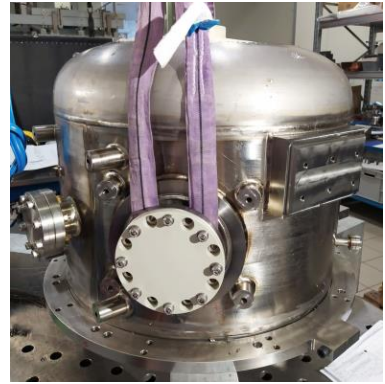

Figure 9: SSR2 Jacketed Cavity.

SSR2 cavities are designed to have a low sensitivity to pressure fluctuation (see Table 1) of the helium bath which, if left unchecked, may result in the largest source of perturbation. The value of $\frac{df}{dp}$ can be extrapolated from step 4 and it is equal to -6 $\frac{Hz}{mbar}$ well within the range required.

## CONCLUSION

The first ppSSR2 jacketed cavity is manufactured and others are soon to be completed. The cavity is within the frequency target range and passed the pressure test.

Table 3: Frequency Shifts during Pressure Test. Values of Frequency are in kHz. Pressure is in bar-g

| Step | Pressure | Freq. |
|------|----------|-------|
| 1    | 0        | -     |
| 2    | 0.5      | -2    |
| 3    | 0        | 1     |
| 4    | 1        | -6    |
| 5    | 0        | 0     |
| 6    | 1.5      | -8    |
| 7    | 0        | 0     |
| 8    | 2        | -11   |
| 9    | 0        | 1     |
| 10   | 2.4      | -14   |
| 11   | 0        | 0     |